# THRESHOLD DETECTION SCHEME BASED ON PARAMETRIC DISTRIBUTION FITTING FOR OPTICAL FIBER CHANNELS

Mohammed Usman*[1] (SMIEEE), Mohd Wajid[2] (SMIEEE), M.Z. Shamim[3], Mohd Dilshad Ansari[4] and Vinit Kumar Gunjan[5]

[1,3]*Department of Electrical Engineering, King Khalid University, Abha-61411, Saudi Arabia*

[2]*Department of Electronics Engineering, Z.H.C.E.T, Aligarh Muslim University, Aligarh-202002, India*

[4]*Department of Computer Science and Engineering, CMR College of Engineering & Technology, Hyderabad 501401, India*

[5]*Department of Computer Science and Engineering, CMR Institute of Technology, Hyderabad 501401, India*

E-mail: musman@ieee.org[1], wajidiitd@ieee.org[2], mzmohammad@kku.edu.sa[3], m.dilshadcse@gmail.com[4], vinitkumargunjan@gmail.com[5]

**Abstract** – When data is transmitted by encoding it in the amplitude of the transmitted signal, such as in Amplitude Shift Keying, the receiver makes a decision on the transmitted data by observing the amplitude of the received signal. Depending on the level of the received signal, the receiver has to decode the received data, with minimum probability of error. For binary transmission, this is achieved by choosing a decision threshold. A threshold detection mechanism is evaluated in this paper, which uses parametric probability distribution fitting to the received signal. Based on our earlier work on visible light communication, the idea is extended to a fiber optic channel and it is found that in a fiber optic channel, the threshold value obtained by fitting a Rayleigh distribution to the received data results in error probability approaching zero.

## 1. INTRODUCTION

Mobile wireless technologies and their applications have continued to evolve and grow at a rapid pace. The demand for high data rates and the requirement of high spectral efficiency has led to the development of several new technologies, which are not just competing but also complementing each other. In systems where information is encoded in the amplitude of the transmitted signal using modulation schemes such as Amplitude Shift Keying (ASK), On-Off-Keying (OOK) and quadrature amplitude modulation (QAM), the demodulator is required to detect the amplitude and make a decision on the received symbol, by comparing the detected amplitude level with some threshold(s). The threshold is usually chosen based on the separation between the amplitude levels representing each bit (0 and 1) and the apriori probability of each bit, $p_0$ and $p_1$ respectively. If the zeros and ones are equally likely, i.e. $p_0 = p_1 = 0.5$, the optimum threshold value lies midway between the two voltage levels. If the apriori probabilities of 0 and 1 are not equal, then the optimum threshold value is closer to the voltage level representing the symbol whose apriori probability is smaller. The receiver therefore needs to know the voltage level used to represent each symbol, the voltage separation between the different symbols as well as their apriori probabilities, in order to choose the optimum threshold, in the sense that it minimizes the probability of error. A threshold detection mechanism based on parametric distribution fitting, in which no prior information is required at the receiver is found to work well over optical wireless channels using visible light communication (VLC), when there is a dominant line of sight (LoS) path between the transmitter and receiver. In the VLC system, an appropriate threshold is obtained by fitting the Rician distribution to the signal level sensed by the photo-receiver [1]. In this paper, threshold detection based on parametric distribution fitting is evaluated for communication over a fiber optic channel. Since there are significant differences between the propagation characteristics of a fiber optic channel and optical wireless channel, the distribution which best models the received signal over a fiber optic channel is different from the Rician distribution. The received signal level at the receiver depends on several factors, such as the transmitted signal level, channel attenuation and fading. For digital communication using Amplitude Shift Keying (ASK), the detection rule at the receiver is:

$$\hat{x} = \begin{cases} 1 \,; y \geq \tau \\ 0 \,; y < \tau \end{cases} \qquad (1)$$

where, $\hat{x}$ is the detected value of the transmitted bit '$x$', '$y$' is the received signal level and $\tau$ is a threshold based on which $\hat{x}$ is obtained. There will be no errors if the detected bit equals the transmitted bit i.e. $\hat{x} = x$. In practical communication systems, due to channel impairments and signal losses, errors occur in the detected bits and it is necessary to minimize the probability of error at the receiver. The received signal level varies as a function of the amount of separation between transmitter and receiver due to signal attenuation and fading due to multipath propagation, which causes a variation in the received signal. In such a scenario, the decision threshold needs to be dynamically adapted in order to minimize error probability. In [2, 3], an adaptive bit detection threshold is calculated by taking the mean of the signal level corresponding to the reception of '1' and '0' respectively. Such a scheme is works well if the apriori probability of transmitting 1 and 0 is equal. The method proposed in this paper for estimating the decision threshold is found to yield error probability approaching zero for both equal and unequal apriori probabilities. In [4], an adaptive threshold detection mechanism is shown to outperform fixed threshold detection and in [5], a threshold detection scheme based on Gaussian approximation has been proposed for optical code division multiple access (O-CDMA) systems. An adaptive log-likelihood ratio (LRT) test threshold detection based on Kalman filtering has been presented in [6]. In this paper, we present a method based on 'distribution fitting' using empirical data, to calculate the threshold which can be dynamically adapted without apriori information. The rest of the paper is organized as follows. In section 2, the threshold estimation procedure based on Rayleigh distribution fitting is discussed. The characteristics of the light source (transmitter) and photo-sensor (receiver) are described in section 3.In section 4, the characteristics of fiber optic channel are presented. Discussion of results is presented in section 5 and conclusions and future work are discussed in section 6.

## 2. THRESHOLD ESTIMATION

In a fiber optic communication system, the signal strength at the receiver depends on the wavelength and brightness of the transmitter, length of fiber optic channel and wavelength of maximum sensitivity of the receiver. The photo-detector converts the received optical signal into a corresponding electrical signal, which is its photocurrent which results in a proportionate photo-voltage 'y'. The probability distribution of 'y' is affected when any of the above mentioned parameters are altered. It is therefore necessary to adapt the decision threshold $\tau$ in order to minimize decision errors. Although a fiber optic channel is immune to external interference, it does suffer from attenuation and time dispersion. The main cause of attenuation is scattering which also results in time dispersion. The scattering characteristic of a light wave propagating through an optical fiber has a Rayleigh distribution [7]. Light traverses a fiber by the process of total internal reflection and there is no direct line of sight component. This is in contrast to a VLC channel which has a LoS component. Hence the distribution of 'y' over a fiber optic channel follows a Rayleigh distribution which is a special case of Rician distribution defined as [8].

$$p(y) = \frac{y}{\sigma^2} exp\left(\frac{-(y^2+s^2)}{2\sigma^2}\right) I_0\left(\frac{ys}{\sigma^2}\right) \qquad (2)$$

Where $I_0(.)$ is the modified Bessel function of the first kind and order zero, and 's' is the non-centrality parameter which arises due to the presence of dominant LOS multipath component and $s \geq 0$. In the absence of a dominant LOS component, as in fiber optic channel, the non-centrality parameter 's' for the Rician distribution equals zero and Rician distribution reduces to a Rayleigh distribution defined as

$$p(y) = \frac{y}{\sigma^2} exp\left(\frac{-y^2}{2\sigma^2}\right) \qquad (3)$$

Where, $\sigma$ is the scale parameter of the Rayleigh distribution. In this paper, we fit distributions to the vector 'y' and the Rayleigh distribution is found to provide the best fit. The decision threshold for binary decision is obtained using the Rayleigh scale parameter as

$$\tau = \sigma + \epsilon \qquad (4)$$

where, $\epsilon$ is a factor used to optimize the threshold. It is empirically determined that $\tau = \sigma + \epsilon$ is a suitable threshold for binary detection over a fiber optic channel. The scale parameter 'σ' is calculated as the maximum likelihood estimate (MLE) of a Rayleigh distribution that provides the best statistical model for the vector '*y*'. Variations in the received optical signal intensity are incorporated in equation 4 as the value of 'σ' depends on the values of '*y*'. The MLE of Rayleigh scale parameter is given as [9].

$$\sigma = \sqrt{\frac{1}{2n}\sum_{i=1}^{n} y_i^2} \qquad (5)$$

Threshold τ is estimated by first computing 'σ' and adding $\epsilon$ to 'σ'.

## 3. TRANSMITTER AND RECEIVER CHARACTERISTICS

The light source used in this work, which acts as transmitter is a plastic fiber optic (PFO) red LED IF-E96. It has a narrow spectral range from 650 – 670 nm, with peak output typically around 660 nm [10]. This is one of the optimum transmission bands for plastic optical fiber used in this work as its attenuation is lowest for wavelengths around 650 nm. The light source is controlled using a micro-controller interfaced switch and is modulated using ASK. The transmitted data is represented by different intensity level of the LED which is controlled by adjusting the duty cycle of a pulse width modulation (PWM) signal, which is also used to control the level separation between '1' and '0'. The photo-receiver used in this work is a LPT80A photo-transistor. While photo-diodes have a faster response time (ns) as compared to photo-transistors (μs), the choice of photo-transistor over photo-diode is made due to its higher gain which simplifies the receiver design. The higher gain of a photo-transistor obviates the need for an amplifier in the receiver. It is a compromise between speed and complexity. The LPT80A has an operational wavelength range from 430 nm to 1070 nm with maximum sensitivity at a wavelength of 850 nm. The relative sensitivity of the photo-receiver at 650 - 660 nm is around 60% of its peak sensitivity [11]. The photo-receiver is also interfaced using a microcontroller to detect its photo-voltage '*y*', based on which the received bits are detected.

## 4. FIBER OPTIC CHANNEL CHARACTERISTICS

Fiber core is usually made of glass or plastic and classified as single mode (<10 μm core diameter) or multimode (25 - 2000 μm) fibers. The fiber optic cable used is a graded index plastic core multimode fiber having a diameter of 1000 μm. The attenuation characteristics of the fiber used in this work is such that 650 nm is an optimum transmission band [12]. Attenuation and time dispersion are less severe in multimode graded index fiber as compared to single mode step index or multimode step index fibers [6].

The attenuation of this fiber is relatively high compared to the state of the art and hence it is suitable only for low rate, short distance applications. It is still chosen due to its low cost and acceptable performance, which suffice the purposes of this study. A fiber optic channel is immune to electromagnetic and radio frequency interference. However, it suffers from attenuation, which is the decay of light as it propagates through the fiber core and represented in terms of dB/km. In contrast, a VLC channel suffers from multipath dispersion due to multiple reflections from objects in the vicinity of the transmitter and receiver as well ambient light, in addition to attenuation [13]. Fiber attenuation is of two types - intrinsic and extrinsic. Intrinsic attenuation is caused due to impurities or imperfections in the fiber manufacturing process. Extrinsic attenuation is caused mainly due to bends and splices (joints) in the fiber. In this work, a single short, straight length of fiber is used and hence extrinsic attenuation is considered negligible. Intrinsic attenuation comprises of two components - absorption of light by the impurity present in the fiber and scattering of light due to collisions between the light particles and molecules of the fiber material - called as Rayleigh scattering [6]. When light scatters, depending on the angle of incidence the scattered path may propagate forward through the fiber core, in which case there is no attenuation or refract out of the fiber core, causing attenuation. Attenuation due to absorption contributes less than 5% of the total attenuation while the bulk of attenuation (more than 95%) is caused due to scattering. Scattering and hence attenuation also depends on the wavelength of light traveling through the fiber. Another form of distortion in an optical fiber is chromatic dispersion. The effects of chromatic dispersion are more severe at higher data rates (when the transmitted pulses are narrower) and result in high error rate[6]. In evaluating the proposed threshold detection mechanism in this work, a narrowband light source is used with relatively lower data rate over a short length of optical fiber (2 m). Hence, chromatic dispersion can also be considered negligible.

## 5. RESULTS AND DISCUSSION

In this paper, threshold estimation based on parametric distribution fitting for binary transmission over fiber optic channel is investigated. Light traverses a fiber optic channel by total internal reflection and the received signal has no direct LoS component. The received signal is therefore expected to follow Rayleigh distribution. Distribution fitting based on Rayleigh distribution is applied to the vector of detected signal levels using maximum likelihood (ML) technique. The ML technique maximizes the likelihood of the distribution parameters to model the observed data [14, 15], which are the detected levels by the photo-receiver. Bayesian information criterion [16] is used to determine the goodness of fit of the hypothesized probability distribution. A detailed analytical description of distribution fitting is described in [1]. The scale parameter 'σ' of the corresponding Rayleigh distribution is used to calculate the decision threshold as described in section 2. The channel impairments taken into consideration are attenuation and Rayleigh scattering.

Results are evaluated for different amounts of separation between the intensity levels used to represent '0' and '1' and for different apriori probabilities $p_0$ and $p_1$ corresponding to '0' and '1' respectively. It is shown that the threshold obtained using the method described in this work yields error probability approaching zero, when the light intensity levels representing '0' and

'1' are spaced sufficiently apart. Figure 1 shows the histograms of the received signal '*y*' for equiprobable 0's and 1's with a separation of 50 between the intensity levels for '0' and '1' and threshold corresponding to $\epsilon = 0$ (1a) $\epsilon = 10$ (1b), $\epsilon = 15$ (1c) and $\epsilon = 20$ (1d). It is noted from figure 1 that the threshold obtained by parametric distribution fitting can be fine-tuned by adjusting the value of $\epsilon$ in order to minimize the error probability. The effect of $\epsilon$ on error probability is more pronounced when the level separation is small.

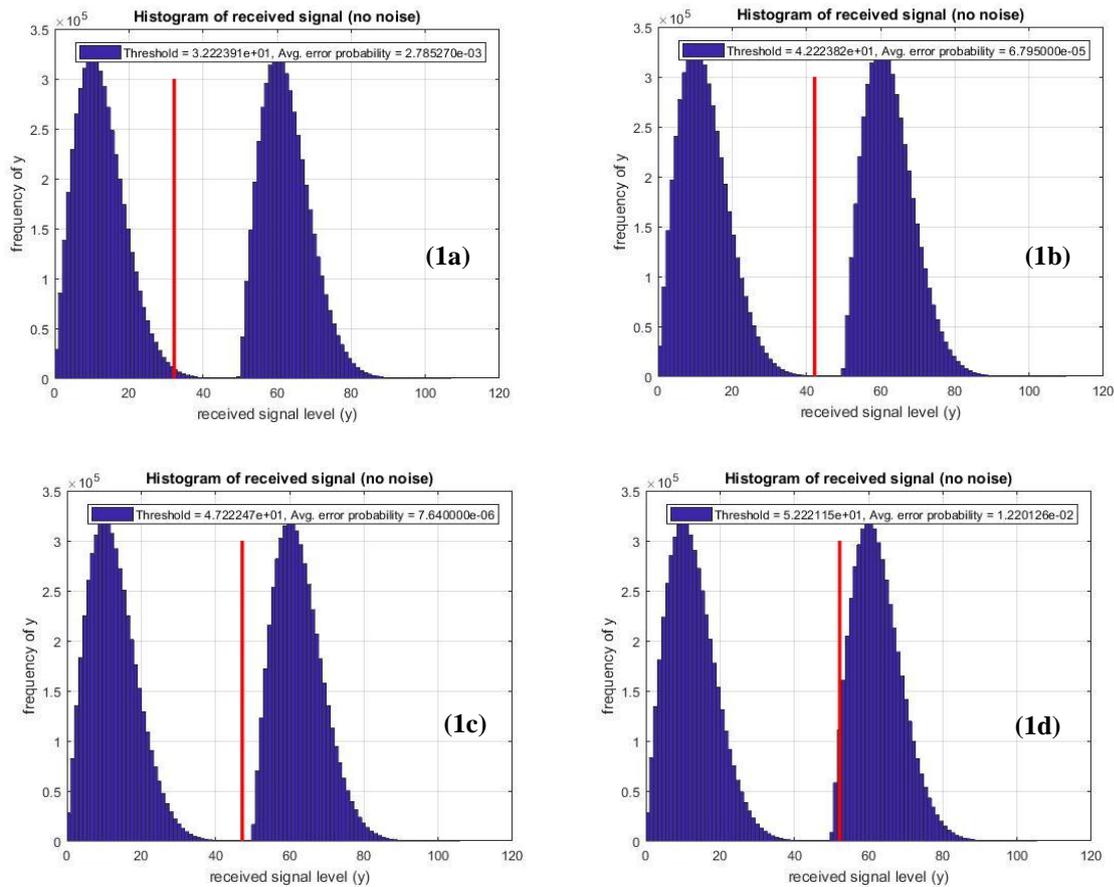

**Fig. 1** Histogram of received signal for $p_0=p_1=0.5$, level separation = 50 with error probability for different thresholds τ. (1a) $\epsilon = 0$, (1b) $\epsilon = 10$, (1c) $\epsilon = 15$ and (1d) $\epsilon = 20$

When the level separation between 0's and 1's is increased, the effect of the value of $\epsilon$ is relatively small on the error probability as can be observed from figure 2 in which the level separation is equal to 70. In figure 3, the histograms and threshold corresponding to unequal apriori probabilities $p_0$ and $p_1$ are shown and it can be observed that the threshold obtained using the proposed method results in small bit error probability. Figures 3 (a), (b) and (c) represent the case for $p_0 = 0.1$ and $p_1 = 0.9$ with $\epsilon = 0$, 5 and 10 respectively. Figures 3 (d), (e) and (f) represent the case for $p_0 = 0.7$ and $p_1 = 0.3$ with $\epsilon = 0$, 10 and 20 respectively. From figures 1, 2 and 3, it can be concluded that the proposed threshold detection method can be used for 'equal' as well as 'unequal' apriori probabilities of the transmitted symbols. By choosing an appropriate value of $\epsilon$, the estimated threshold results in small bit error probability in each case.

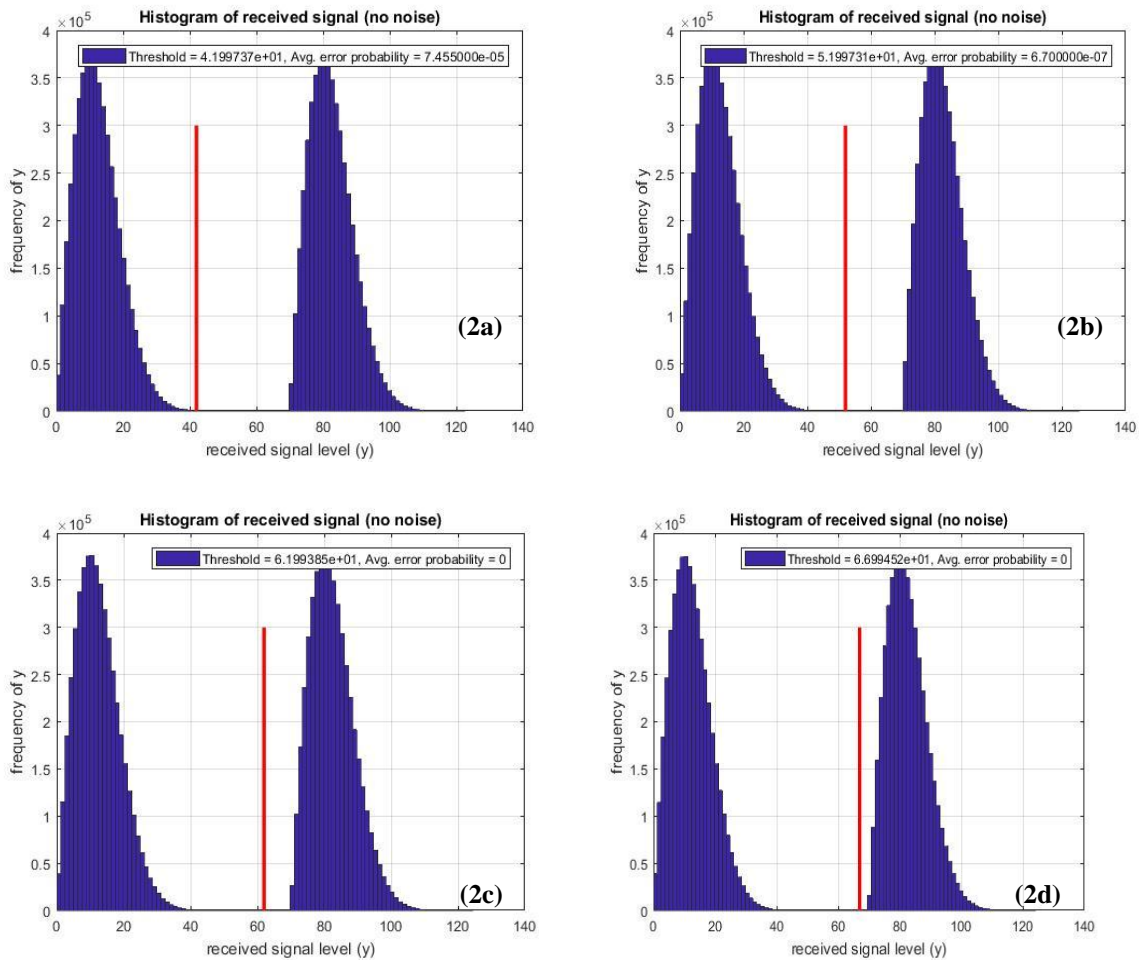

**Fig. 2** Histogram of received signal for $p_0=p_1=0.5$, level separation = 70 with error probability for different thresholds τ. (2a) $\epsilon = 0$, (2b) $\epsilon = 10$, (2c) $\epsilon = 20$ and (2d) $\epsilon = 25$

The variation of BER as a function of level separation between 0 and 1 for three different apriori probabilities of the input binary data is shown in figure 4. While the threshold obtained using distribution fitting of the received signal level yields small BER for all combinations of apriori probabilities, the BER values, for a given level separation, are relatively lower when the binary 1's are more likely than the binary 0's. This is attributed to the fact that 1's are represented using higher intensity level than 0's, resulting in smaller error rate when there are more 1's. When 0's are more likely than 1's, it results in lower average signal to noise ratio since 0's are represented using lower intensity level, resulting in higher error rate.

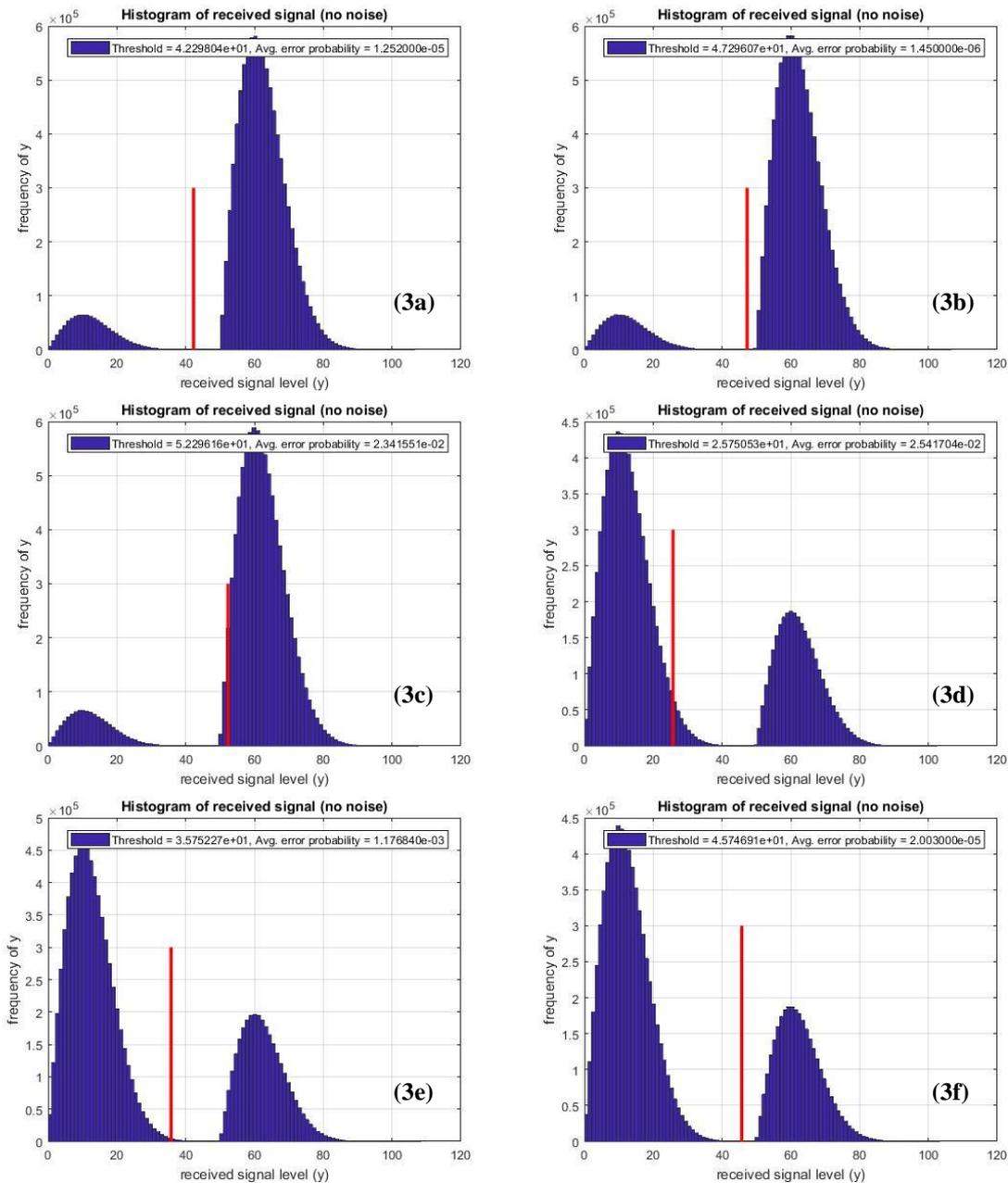

**Fig. 3** Histogram of received signal for level separation = 50 (3a) $p_0$= 0.1, $\epsilon$ = 0, (3b) $p_0$= 0.1, $\epsilon$ = 5, (3c) $p_0$= 0.1, $\epsilon$ = 10 and (3d) $p_0$= 0.7, $\epsilon$ = 0, (3e) $p_0$= 0.7, $\epsilon$ = 10, (3f) $p_0$= 0.7, $\epsilon$ = 20, with error probability for different thresholds $\tau$

It must be noted here that the sources of noise taken into consideration are transmitter circuitry, receiver circuitry, transmitter LED, photo-receiver, and any imperfections in the optical fiber. The effects of external noise and extrinsic optical interference are not considered in this work.

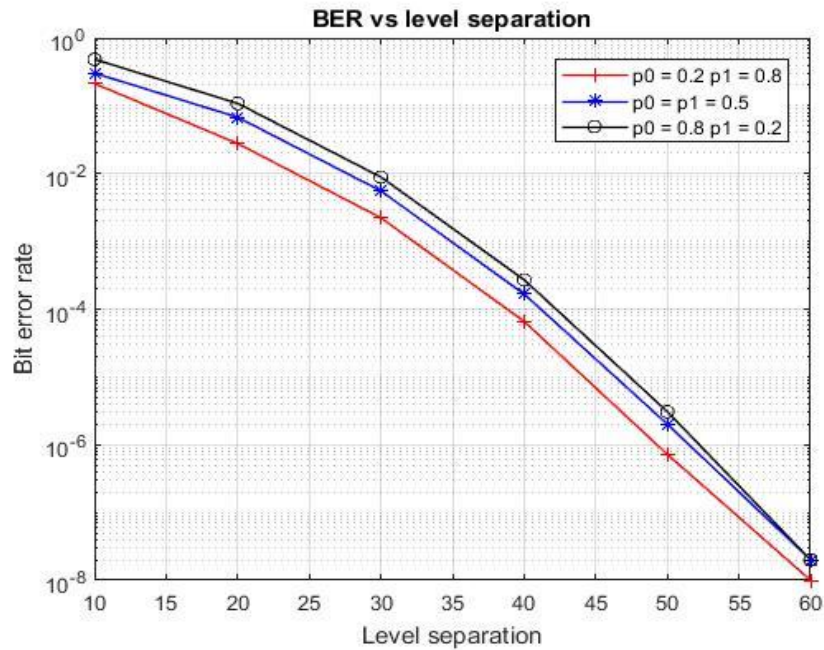

**Fig. 4** BER vs level separation for different apriori probabilities

## 6. CONCLUSIONS AND FUTURE WORK

In this paper, a binary decision threshold calculation method based on fitting parametric distributions to the observed data, which was proposed in our earlier work [1] for visible light communication is adapted for communication over optical fiber. While in the VLC system, Rician distribution provided a suitable threshold, in fiber optic system, it is found that Rayleigh distribution provides the best fit to the received data and threshold is obtained based on the Rayleigh scale parameter 'σ' as defined in equation 4. The binary decision based on the obtained threshold results in small BER that approaches zero. When the level separation between the intensity levels representing 0's and 1's is sufficiently large, the Rayleigh scale parameter requires little adjustment. However, when the level separation is small, the scale parameter 'σ' requires adjustment which is achieved by adding a parameter ϵ to 'σ'. It is found that the proposed threshold estimation method works well for both equal as well as unequal apriori probabilities of the binary symbols. The BER is marginally higher when the likelihood of 0's being transmitted is more, for a given level separation. The proposed method of threshold detection is adaptive when the apriori probabilities for 1's and 0's change. In future, it is planned to investigate the feasibility of the proposed method when external noise and interference are also included in the distortion introduced in the propagation medium. Further, threshold detection based on image processing applied to histogram images may be investigated using techniques similar to those in [17].

Declaration of interest: None

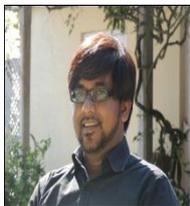
MOHAMMED USMAN received B.E in Electronics and Communications from Madras University, India in 2002. He was awarded M.Sc and PhD from University of Strathclyde, United Kingdom in 2003 and 2008 respectively. He is the recipient of the University scholarship from Strathclyde for his PhD. He has more than a decade of experience in academics and academic administration. He is a senior member of IEEE, USA and IEEE Communications Society (ComSoc). He has been TPC chair and organizing chair for IEEE conferences and actively involved in IEEE activities. He received the best faculty award from the College of Engineering at King Khalid University and also received the award for best project, both in 2016. His research is focussed on technologies for next generation communication systems, probabilistic modelling and biomedical speech processing.

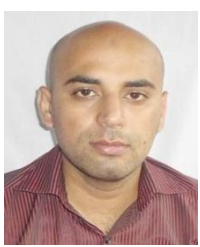
MOHD WAJID is an assistant professor in the Department of Electronics Engineering, AMU, Aligarh. Wajid did his B.Tech (Electronics) from AMU - Aligarh and M.Tech (VLSI and Embedded Systems) from IIIT Hyderabad. He obtained PhD degree in Signal Processing from IndianInstitute of

Technology Delhi, India. Before joining Aligarh Muslim University, he was associated with Jaypee University of Information Technology, Texas Instruments, Xilinx India Technology Services Private Limited, and Blue Star Limited. He is also a Senior Member of IEEE, USA.

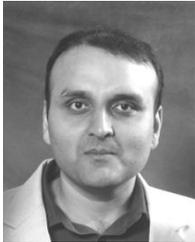

MOHAMMED ZUBAIR M. SHAMIM received his M.Eng. degree in electronics and electrical engineering from the University of Dundee, Dundee, United Kingdom in 2003. Further he received his Ph.D. degree in electronics engineering and applied physics from the same university in 2008. From 2005 to 2009 he was a Research and Development Engineer with Quantum Filament Technologies Ltd, United Kingdom. From 2009 to 2011 he was a Senior Research Scientist at the University of Stuttgart, Germany. His current research interests include fabrication of functional micro-nanostructured surfaces using pulsed excimer laser crystallization for different applications and GPGPU accelerated computing for deep learning.

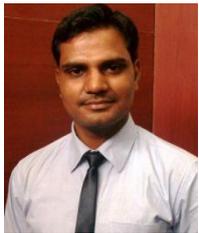

MOHD DILSHAD ANSARI is currently working as Assistant Professor in the department of Computer Science & Engineering at CMR College of Engineering & Technology, Hyderabad, India. He received his Ph.D. in 2018 in Image Processing from Jaypee University of Information Technology, Waknaghat, Solan, HP, India. He obtained his M.Tech in Computer Science and Engineering in 2011 and B.Tech in Information Technology from Uttar Pradesh Technical University, Lucknow, UP in 2009. He has published more than 20 papers in International Journals and conferences. He is the Member of various technical/professional societies such as IEEE, UACEE and IACSIT. He has been appointed as Editorial/Reviewer Board and Technical Programme Committee member in various reputed Journals/Conferences. His research interest includes Digital & Fuzzy Image Processing, IOT, and Cloud Computing.

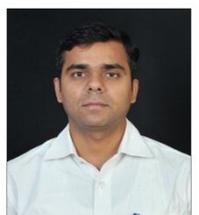

Vinit Kumar Gunjan is Associate Professor in Department of Computer Science & Engineering at CMR Institute of Technology Hyderabad (Affiliated to Jawaharlal Nehru Technological University, Hyderabad). An active researcher; published research papers in IEEE, Elsevier & Springer Conferences, authored several books and edited volumes of Springer series, most of which are indexed in SCOPUS database. He was awarded with the prestigious Early Career Research Award in the year 2016 by Science Engineering Research Board, Department of

Science & Technology Government of India. He is a Senior Member of IEEE and an active Volunteer of IEEE Hyderabad section; volunteered in the capacity of Treasurer, secretary & Chairman of IEEE Young Professionals Affinity Group & IEEE Computer Society. He was involved as organizer in many technical & non-technical workshops, seminars & conferences of IEEE & Springer. During the tenure he had the honour of working with top leaders of IEEE and awarded with best IEEE Young Professional award in 2017 by IEEE Hyderabad Section.